\documentclass[a4paper,twocolumn,11pt]{quantumarticle}
\pdfoutput=1
\usepackage[utf8]{inputenc}
\usepackage[english]{babel}
\usepackage[T1]{fontenc}
\usepackage{amsmath}
\PassOptionsToPackage{hyphens}{url}\usepackage{hyperref}

\usepackage{tikz}
\usepackage{amsmath}
\usepackage{amssymb}
\usepackage{amsthm}
\usepackage{fancyhdr}
\usepackage{t1enc}
\usepackage{geometry}
\usepackage{chngcntr}
\usepackage[frozencache,cachedir=.]{minted}
\usepackage{listings}
\lstdefinestyle{python}{
language=Python, 
basicstyle=\ttfamily\scriptsize,
morekeywords={self},              
keywordstyle=\bfseries\color{blue},
frame=tb,                         
showstringspaces=false
}

\usepackage{graphicx}
\usepackage{bbold}
\usepackage{physics}
\usepackage{bm}
\usepackage{changepage}
\usepackage{amsfonts}
\usepackage{booktabs}
\usepackage{caption}
\usepackage[capitalise]{cleveref}
\usepackage{csquotes}
\begin{document}


\title{Variational Quantum Optimization Benchmark Suite for Airline Crew Pairing and More}

\author{D\'avid Sipos}
\affiliation{Hamburg University of Technology, Institute for Algorithms and Complexity, Germany}
\email{david.sipos@tuhh.de}
\orcid{0009-0002-5398-1984}
\author{Andr\'as Cz\'egel}
\affiliation{Institute of Informatics, University of Szeged, 6720 Szeged, Hungary}
\orcid{0009-0007-9156-0124}
\email{czegel@inf.u-szeged.hu}
\author{Bogl\'arka G.-T\'oth}
\affiliation{Institute of Informatics, University of Szeged, 6720 Szeged, Hungary}
\email{boglarka@inf.szte.hu}
\orcid{0000-0002-0927-111X}

\maketitle

\begin{abstract}
We introduce a set of open-source packages that form a highly extensible framework for quantum optimization. One design goal of the system is the inclusion of a command line based configuration system for setting up experiments. The possible options are derived using well-known Python packages and presented to the user intuitively, allowing the configuration of repeatable variational quantum optimization experiments. We give an example of using the system through the Airline Crew Pairing problem, a highly relevant industrial problem, and the MaxCut problem, for which instances of manageable size are readily available.
\end{abstract}

\section{Introduction}
In today's Noisy Intermediate-Scale Quantum (NISQ) era~\cite{nisq} with quantum computers only possessing qubits in the range of hundreds to low-thousands and fully fault-tolerant systems far on the horizon, many have turned to hybrid quantum-classical algorithms~\cite{Bharti2022} where part of the computation is offloaded to classical computers.

One family of such hybrid algorithms are the Variational Quantum Algorithms (VQA), introduced with the Variational Quantum Eigensolver (VQE)~\cite{vqa} and later the Quantum Approximate Optimization Algorithm (QAOA)~\cite{qaoa} and its many variants~\cite{qaoa_review}.
Such hybrid algorithms have shown promising results among many fields in machine learning~\cite{qml_Havlek2019, Shingu2021}, quantum-chemistry and physics~\cite{qchem_Smart2021, qchem_Gao2021, Jones2019, Delgado2021}.

To solve a problem using VQAs we need to define a cost function that we can map to a quantum circuit. For any single problem, there can be many different circuit designs, which are not necessarily problem-specific. These circuit designs are referred to as \emph{ans\"atze} (singular ansatz), ``educated guesses'' about the best way to map the problem to a quantum circuit and, in a way, are akin to the architectural choices in building a neural network. Different ans\"atze can provide different benefits, such as reducing the Hilbert space explored by the optimizer or altering it such that infeasible solutions cannot appear as measurement outcomes. Other ans\"atze may try to reduce the depth of the quantum circuit or adapt to the hardware of the quantum computer running the subroutine.

Over the years, VQAs have received extensive attention in the quantum optimization literature, producing several algorithm variants~\cite{Abbas2024}. Similarly, many advances have been made on the hardware side, allowing experiments to be conducted on ever-growing scales using architecturally diverse quantum computing systems through accessible cloud-based computing providers~\cite{ibm, braket}. What we have seen less of, partly due to the size of the problems involved, are experiments and software solutions aimed at more industrial applications. That gap is what we aim to fill with our framework.

\medskip
In the airline industry, a significant source of operating costs is labor costs~\cite{a4a_paci,a4a_state} associated with the operation of aircraft. As such, significant resources are invested to optimize the costs of airline crews while complying with different labor regulations. Due to the complex nature of crew planning, the problem is often considered as a sequence of subproblems, each with its own set of challenges and possibilities for optimization.

\newpage
In the \textsc{Airline Crew Pairing} (ACP) subproblem, a sequence of flights is called \emph{pairing} if the home base from which the first flight departs is the same as the one at which the last flight arrives, and it meets regulatory requirements and constraints placed by airlines~\cite{deveci2018evolutionary, muter2013solving, borndorfer2006column}.

\medskip
In \cref{sec:framework}, we go into detail about the motivation behind the optimization framework and give an overview of the architecture of the software suite. In\cref{sec:acp}, we introduce the ACP problem as an example of using the framework as an end-to-end optimization suite to handle industrial problems. We describe how the problem was modeled in a way that allows it to be solved using the framework. This serves as a guide for introducing other optimization problems. We then conclude with a set of experimental results on manageable problem instances of ACP and \textsc{MaxCut} in \cref{sec:results} and summarize our work in~\cref{sec:summary}.

\section{Hybrid optimization framework}\label{sec:framework}

\subsection{Motivation}\label{sec:motivation}

The purpose of the optimization suite\footnote{\url{https://github.com/david-sipos/vqaopt}} is to provide an end-to-end solution to evaluate the applicability of variational quantum algorithms to industrial problems. The software itself can be used, without modification, as an off-the-shelf solver while allowing the alteration of each part of the optimization process so that new advances in quantum optimization can be easily tested and integrated into existing processes.

The system itself is designed to be easy to use in several settings. The core of the system is a well-documented set of definitions describing parts of the optimization process, which can be used to create Python programs and notebooks for quick experiments.

When used as an off-the-shelf solver, the system is easily configured via a human-readable configuration file. The creation of configuration files is supported by a CLI tool that prompts the user to describe the problem to be solved and the set of steps to be taken to solve it. This allows the system to be used without touching the system's source code or writing Python scripts.

\subsection{Plugin system}\label{plugins}
Plugins provide functionality as distinct pieces of the optimization process.
They are Python classes that implement plugin interfaces by inheriting from a plugin base class.
Plugins are registered upon importing their definition being encountered by the Python interpreter, which allows them to be loaded automatically.
This allows third parties to contribute plugins by contributing entry points to the appropriate group. This makes using plugins as easy as installing them through a package manager (e.g. \emph{pip}~\cite{pip}).
Additionally, plugins can be registered by specifying a list of Python filenames from which to load plugins in the configuration file or as arguments for the CLI tool.

\subsubsection{Problem Loaders}
The optimization process starts with a problem loader. The purpose of such a plugin is to load or generate data and formulate the instance, or instances, of the problem to be solved. The problem is represented by an object that may store arbitrary information about the problem instance alongside the name of the problem and a collection of its \emph{forms}, instances of different problems that are equivalent to the original problem.

\subsubsection{Platforms}
The optimization suite supports multiple quantum computing providers by allowing users to implement plugins to interface with them, such as Qiskit\footnote{\url{https://github.com/david-sipos/vqaopt-platform-qiskit}} and Braket\footnote{\url{https://github.com/david-sipos/vqaopt-platform-braket}}.
Each platform plugin must define a set of methods needed for the execution of VQAs, but may include arbitrary functionality to utilize the unique offerings of different quantum computing services.

The optimization suite uses the circuit model to describe quantum algorithms and requires that platform integrations define a mapping to a common set of gates such that VQAs can be defined platform independently.

\subsubsection{Ans\"atze}
The quantum circuit executed with the help of the computing provider is determined by the ansatz of choice and constructed by the corresponding ansatz plugin. The plugin makes use of the quantum gates expected to be implemented on the target platform and allows for the definition of variational circuits using an intuitive API, for a simple example see \cref{lst:ansatz}. The suite contains many implemented options, for their description and comparison, see \cref{appendix:ansatze}.

\begin{lstlisting}[style=python, caption=Defining an ansatz,label=lst:ansatz, columns=fullflexible] 
def ansatz(self,
  qubits: int,
  depth: int,
  problem: IsingProblem,
) -> None:
  
  gamma = Parametric("gamma")
  beta = Parametric("beta", upper=np.pi)

  for j in range(qubits):
    gates.h(j)

  for i in range(depth):
    for j_prime in range(qubits):
      for j in range(j_prime):
        gates.rzz(
          2 * gamma[i] * problem.J[j, j_prime], 
          j,
          j_prime,
        )
        for j in range(qubits):
          gates.rz(2 * gamma[i] * problem.h[j], j)
        for j in range(qubits):
          gates.rx(2 * beta[i], j)
\end{lstlisting}

\subsubsection{Reductions}
Problem instances must be in a form appropriate for the construction of the ansatz used. To facilitate this, it is possible to define reductions between problems via reduction plugins. Such a plugin implements a one-way reduction from a problem instance to an equivalent instance of a different problem. These reductions define a directed graph from which the shortest path is chosen to convert the problem loaded from the dataset to a problem required by the ansatz.

We hope that this system will free up time that users would spend modeling their problems in perhaps less intuitive forms. For example, when solving the ACP problem, there is no need to manually write the problem as an Ising Hamiltonian, existing reductions will handle that.

\subsubsection{Initializers}
Multiple works~\cite{Galda2021, Lee2021} suggest that the use of different initialization strategies in the optimization process can yield improvements in the quality of the solutions. As such, the optimization suite allows for implementing initialization strategies.

\subsubsection{Optimizers}
To evaluate different global optimization algorithms, the suite can integrate them using optimizer plugins. Such a plugin, when given a black-box function and initial parameters obtained from an initializer plugin, adjusts the parameters until an optimum is found or some other termination criterion is met. A more detailed description of currently implemented optimizers in the suite can be found in \cref{appendix:opt}.

\subsubsection{Result processors}
To interpret the results of the optimization process, result processor plugins can be defined. These plugins receive information about the entirety of the optimization process and are meant to generate different logs and plots in two rounds.
First, the result processors can extract data every time a solution is found (the optimizer terminates).
This data is then collected across multiple problems (or multiple attempts at solving a problem) and passed back to the result processors to generate output files from the aggregated data.
\subsection{Command line interface}
One goal of the framework is to provide an easy way to configure and execute experiments through a command line interface, thus requiring little to no interaction with or knowledge of the underlying source code.
Upon installation of the package, the command line utility \texttt{vqaopt} is installed which allows the execution of two command groups through the terminal.

The \texttt{config} command group contains commands to configure experiments, while the \texttt{run} command group is for executed experiments and related tasks.
Similarly to plugins, commands are registered using entry points and depending on their definition are placed in the appropriate group.

\subsubsection{Plugin configuration}
Every plugin in the framework is based on the Pydantic~\cite{pydantic} data validation library.
This allows for an easy and robust solution for describing how a plugin can be configured and attach metadata such as a user-friendly name, example values, or data validation.

Our framework provides a way, through the \texttt{Configurator} class, to interpret the information attached to the configurable fields of plugins, and present them as freeform text or selection input in a series of configuration menus, such as the one in \cref{lst:config}.

The configuration can then be saved to file and used by the command line tools to instantiate the configured plugins, by passing the configuration to the \texttt{instantiate} function, and to orchestrate experiments.

One such command line tool is \texttt{QAOAExperiment}, which configures an optimization pipeline that starts by loading a problem instance (or set of instances) and solves it using a variational quantum algorithm, defined by the optimizer, ansatz, and initialization strategy.
The results are then collected and processed using result processor plugins.

\lstset{
    literate={»}{{\guillemotright\:\:}}1
}

\begin{lstlisting}[basicstyle=\ttfamily\scriptsize,caption=Configuring the Qiskit Platform,label=lst:config, columns=fullflexible,frame=tb]
? Select a field to configure:
(Use arrow keys)
   D Run using sim
 » D Path to .env file: None
   D Backend name:
   D Simulation method: statevector
   D Floating point precision: double
   D Maximum number of jobs: None
   D Maximum number of shots: None
   D Enable truncation: True
\end{lstlisting}

\section{Solving Airline Crew Pairing using the suite}\label{sec:acp}

Based on the architectural description, it is clear that the suite can be turned into a generic solver for any problem with the suitable plugins defined. Since the current capabilities of quantum optimization algorithms are restricted by the hardware, most use cases revolve around small problem instances.
Considering the potential future industrial applications of quantum computers, however, we decided to implement the ACP problem with appropriate example plugins and data sets. Its significance in the industry and complexity are already discussed in the introduction, so let us now dive into how the problem is solved using the suite.
To do this, we describe how the problem can be formulated in a way such that it is solvable using a VQA. This also allows us to give a brief description of some other problems currently implemented.

\subsection{Airline Crew Pairing}
Here we remind the reader of a couple of terms used when discussing the ACP problem.
\begin{itemize}
    \item \emph{Flight leg} A flight from a departure airport to an arrival airport.
    \item \emph{Duty} A valid sequence of a single day's worth of flight legs. A sequence of legs is valid if it conforms to the considered regulations and additional constraints some of which are discussed later in Appendix \eqref{rules}.
    \item \emph{Pairing} A valid sequence of duties centered around a home base. As duties themselves are sequences of legs, a pairing can be thought of as a sequence of flight legs that depart from and ultimately arrive at the same airport. A sequence of duties is a valid pairing if the duties themselves are valid and the sequence conforms to the considered regulations and additional constraints some of which are also discussed in Section \eqref{rules}.
\end{itemize}
Given a set of flight legs alongside pairings with their associated costs, our goal in the ACP problem is to select a subset of the pairings, which covers all flight legs while minimizing the total costs.

\subsection{Minimum Cost Exact Cover}
The \textsc{Exact Cover} problem is concerned with selecting subsets of a set $S$ from a family of subsets $\mathcal{F}$ such that the selected subsets contain each element of $S$ exactly once.

Formally, given a set $S = \{a_1 \dots a_n\}$ and family $\mathcal{F} \subseteq \mathcal{P}(S)$ an \emph{exact cover} of $S$ is a subset $\mathcal{F}^{*}\subseteq \mathcal{F}$ that satisfies the following:
\begin{align}
    s_i \cap s_j &= \emptyset \quad \forall s_i,s_j \in \mathcal{F}^*, i\neq j,\\
    \underset{s_j \in \mathcal{F}^*}{\bigcup} s_j &= S\enspace.
\end{align}

An optimization version of the \textsc{Exact Cover} problem can be formulated as finding an exact cover using a minimum number of subsets.\\
Generalizing this further we get the \textsc{Minimum Cost Exact Cover} (MCEC) problem where each subset $s_i$ has an associated cost $c_i$, and we call an exact cover \emph{minimum} if the sum of the costs of the included subsets, $\sum_{s_j \in \mathcal{F}^*} c_j$, is minimum. The optimization problem where our goal is to use a minimum number of subsets is a special case of the MCEC problem where every subset shares the same positive cost.

It is easy to see that with flight legs being the elements of set $S$, while valid pairings with their associated costs being the subsets in $\mathcal{F}$, the ACP problem can be viewed as MCEC.

\subsection{Quadratic Unconstrained Binary Optimization}
Given a symmetric matrix $Q \in \mathbb{R}^{n \cross n}$, finding a binary vector $x \in \{0,1\}^n$ such that the term $x^\intercal Q x$ is minimized is often referred to as the \textsc{Quadratic Unconstrained Binary Optimization} (QUBO) problem.
It is a combinatorial optimization problem of great importance when solving problems using quantum optimization~\cite{Date2021, McCollum2021, Jun2024} due to its connection to the Ising model, which we will discuss in \cref{phys-sys}.

We can rewrite the ACP problem formulated as MCEC into a QUBO problem by first noticing that the original MCEC formulation is equivalent to the following binary linear optimization problem:
\begin{align}
    \min& \quad z=\sum_{j=1}^{m}c_jx_j&\\
    \textrm{s.t.}& \quad \sum_{j=1}^{m}b_{ij}x_j=1& \quad \forall i \in \{1,\dots,n\}\\
    &\quad x_j \in \{0,1\} &\forall j \in \{1,\dots,m\} ,
\end{align}
where $b_{ij}$ is $1$ if subset $s_j\in \mathcal{F}$ contains element $a_i \in S$, otherwise $0$. The binary variables $x_j$\\$(j=1,\ldots,m)$ determine whether a subset $s_j$ is part of the exact cover ($x_j = 1$), or not ($x_j=0$).

We can turn this into a QUBO problem by placing penalty terms in the objective that fulfill similar roles as the constraints.
For each constraint $i$ we will use the penalty term $\left( 1- \sum_{j=1}^{m}b_{ij}x_j \right)^2$ mentioned by Lucas~\cite{lucas_2014}. It will be minimum ($0$) if constraint $i$ is satisfied. Accordingly, the reformulated objective function is
\begin{equation}\label{eq-qubo}
    \min z=D\sum_{i=1}^{n} \left( 1- \sum_{j=1}^{m}b_{ij}x_j \right)^2 + \sum_{j=1}^{m}c_jx_j,
\end{equation}
where the first term penalizes violating the constraints, and the second term encodes the cost of a particular solution.
Note that the penalty term is scaled by a large enough factor $D$ to ensure that \textquote{it doesn't make sense} to violate a constraint.

\subsection{Ising problem}\label{phys-sys}
We refer to finding the ground state of a system described by the Ising model as the \emph{Ising problem}. Many instances of NP-complete problems can be formulated as instances of the Ising problem~\cite{lucas_2014}.

In the problem, we take an Ising Hamiltonian of the form
\begin{equation}\label{cost_hamiltonian}
    \mathcal{H}_C = \sum_{\substack{j,j'=1 \\ j < j'}}^{m}J_{jj'}\sigma^z_j\sigma^z_{j'} + \sum_{j=1}^{m}h_j\sigma^z_j,
\end{equation}
and try to find the the quantum state $\ket{\phi}$ for which the expected value $\bra{\phi}\mathcal{H}_C\ket{\phi}$ is minimum.
By setting $\forall j,j' \in \{1\ldots m\}$
\begin{align}
    J_{jj'} &= \frac{D}{2}\sum_{i=1}^n b_{ij}b_{ij'} , 
    \\
    h_j &= \frac{D}{2}\sum_{i=1}^n b_{ij}\left( \sum_{j'=1}^m b_{ij'}-2 \right) + \frac{c_j}{2} , 
\end{align}
we can encode the ACP problem as an Ising Hamiltonian.

\subsection{Using a VQA to find the minimum energy state}
To find the ground state of the Hamiltonian~$H_C$, corresponding to an optimal solution to the ACP~problem, we can define a black-box function, which is evaluated using a quantum computer and a classical computer to optimize this function.
In the suite, we can do this by creating an ansatz, which describes the quantum circuit to be sampled to estimate the expected value of the Hamiltonian.

\subsection{ACP implementation module}
We provide an implementation\footnote{\url{https://github.com/david-sipos/optsuite-crew-pairing}} for a module that integrates the crew pairing problem into the framework by implementing the problem, problem loaders, and reductions necessary. The module also defines additional plugins required for the construction of a problem instance, highlighting that one need not stick only to plugins defined in the core package.
After installation, the problem is part of the system, and experiments involving it can be easily configured.

\section{Results}\label{sec:results}
We ran an experiment on a small instance of the ACP problem, constructed in the style described by Kasirzadeh et al.~\cite{dataset_Kasirzadeh2017}.
We used a genetic algorithm to optimize the parameters of the ma-QAOA ansatz.
\begin{figure*}[ht]
  \centering
  \includegraphics[width=\textwidth]{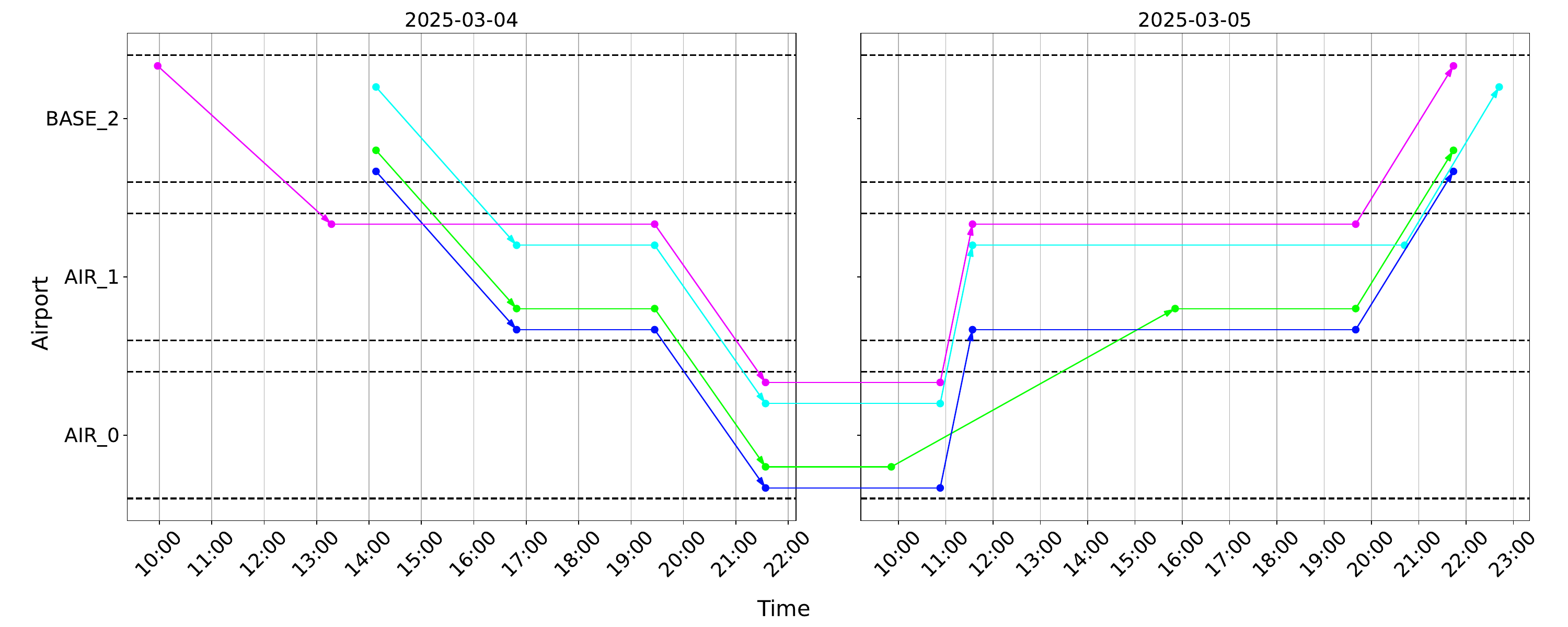}
  \caption{Most likely solution after optimization.}
  \label{fig:pairings}
\end{figure*}
The pairing most likely to be measured can be seen in \cref{fig:pairings}.

\begin{figure*}[ht]
  \centering
  \includegraphics[width=\textwidth]{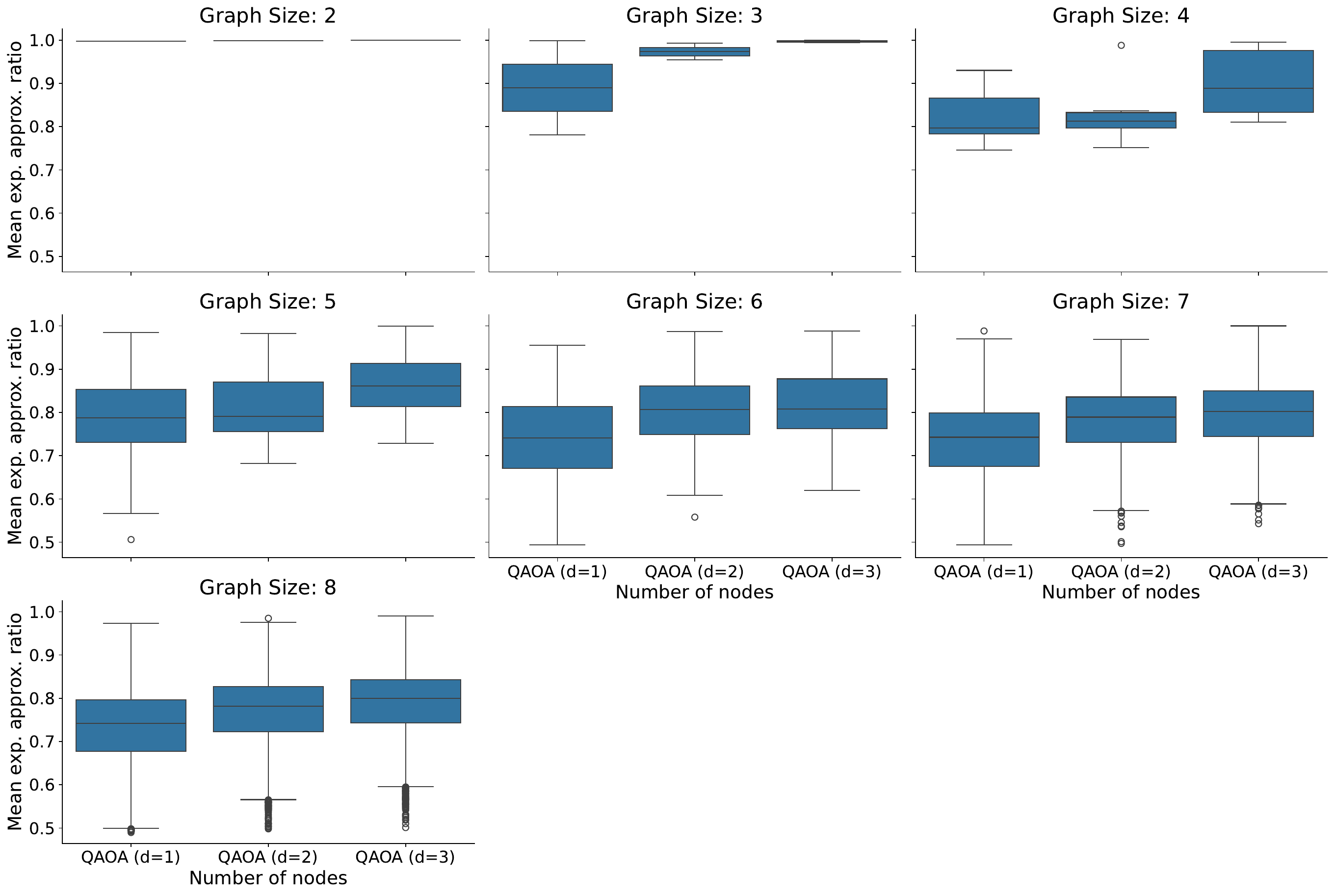}
  \caption{Expected approximation ratios over different problem sizes.}
  \label{fig:approx-box}
\end{figure*}

We also used the suite to compare how the depth of the QAOA ansatz affects the results inspired by the experimental results in the work of Hermann et al.~\cite{https://doi.org/10.48550/arxiv.2109.11455}.
For this, we solved the \textsc{MaxCut} problem on all connected $n$-node graphs with $n\in[2,8]$.
For all instances, the approximation ratio and expected approximation ratio were calculated for each ansatz, depicted in \cref{tab:approx-ratio}, where the approximation ratio is based on the most likely solution, and the expected approximation ratio is the expected value of the approximation ratio based on a sampling of the probability distribution induced by the quantum circuit. \cref{fig:approx-box} shows the expected approximation ratios broken down by instance size.

\begin{table*}[ht]
    \centering
    \captionsetup{width=.8\linewidth}
  \caption{Average approximation ratios and expected approximation ratios for various ans\"atze.}
  \label{tab:approx-ratio}
  \begin{tabular}{cccl}
    \toprule
    & $\text{QAOA}_1$ & $\text{QAOA}_2$ & $\text{QAOA}_3$\\
    \midrule
    Avg. approx. ratio  & 0.8515 & 0.9099 & 0.9269\\
    Avg. exp. approx. ratio & 0.7360 & 0.7726 & 0.7903\\
  \bottomrule
\end{tabular}
\end{table*}

\begin{figure*}[ht]
  \centering
  \includegraphics[width=0.8\textwidth]{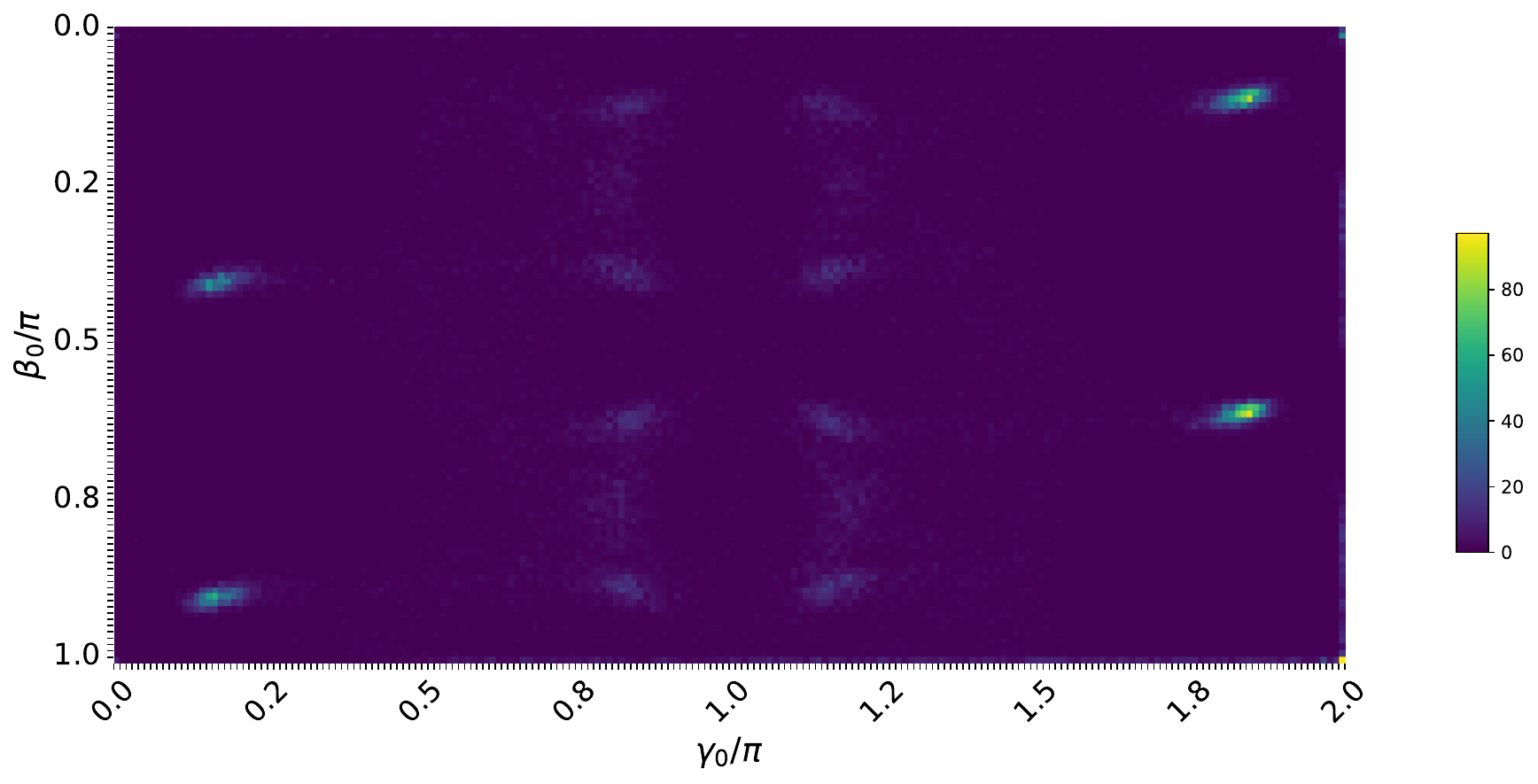}
  \caption{Angle patterns of QAOA with depth 1.}
  \label{fig:angle-1}
\end{figure*}

We plotted the parameters to which the optimizer converged in \cref{fig:angle-1} further confirming that certain parameters are preferred regardless of the problem instance.

\section{Summary}\label{sec:summary}
We introduce a highly modularized suite for quantum optimization. The suite has an industrially relevant problem implemented with several state-of-the-art quantum and classical solver options. 

Furthermore, the suite is tailored for ease of use in many use cases: benchmarking hardware, benchmarking new algorithms, comparison between different hardware-software stacks, and between different solver solutions are just some of the possible applications that we hope can serve as a solid foundation for further experimental work.
As with many software projects, this framework still has a lot of potential for further development.
The introduction of more problems, detailed documentation, and other known algorithms need to be kept in mind to ensure that the project is accessible and up to date.

\bibliography{bibliography.bib}

\begin{thebibliography}{10}

\bibitem{nisq}
John Preskill.
\newblock ``Quantum {C}omputing in the {NISQ} era and beyond''.
\newblock \href{https://dx.doi.org/10.22331/q-2018-08-06-79}{{Quantum} {\bf 2}, 79}~(2018).

\bibitem{Bharti2022}
Kishor Bharti, Alba Cervera-Lierta, Thi~Ha Kyaw, Tobias Haug, Sumner Alperin-Lea, Abhinav Anand, Matthias Degroote, Hermanni Heimonen, Jakob~S. Kottmann, Tim Menke, Wai-Keong Mok, Sukin Sim, Leong-Chuan Kwek, and Alán Aspuru-Guzik.
\newblock ``{Noisy intermediate-scale quantum algorithms}''.
\newblock \href{https://dx.doi.org/10.1103/revmodphys.94.015004}{Reviews of Modern Physics {\bf 94}, 015004}~(2022).

\bibitem{vqa}
Alberto Peruzzo, Jarrod McClean, Peter Shadbolt, Man-Hong Yung, Xiao-Qi Zhou, Peter~J. Love, Alán Aspuru-Guzik, and Jeremy~L. O’Brien.
\newblock ``A variational eigenvalue solver on a photonic quantum processor''.
\newblock \href{https://dx.doi.org/10.1038/ncomms5213}{Nature Communications {\bf 5}, 4213}~(2014).

\bibitem{qaoa}
Edward Farhi, Jeffrey Goldstone, and Sam Gutmann.
\newblock ``A quantum approximate optimization algorithm''~(2014).
\newblock \url{https://arxiv.org/abs/1411.4028}.

\bibitem{qaoa_review}
Kostas Blekos, Dean Brand, Andrea Ceschini, Chiao-Hui Chou, Rui-Hao Li, Komal Pandya, and Alessandro Summer.
\newblock ``A review on quantum approximate optimization algorithm and its variants''.
\newblock \href{https://dx.doi.org/https://doi.org/10.1016/j.physrep.2024.03.002}{Physics Reports {\bf 1068}, 1--66}~(2024).

\bibitem{qml_Havlek2019}
Vojtěch Havlíček, Antonio~D. Córcoles, Kristan Temme, Aram~W. Harrow, Abhinav Kandala, Jerry~M. Chow, and Jay~M. Gambetta.
\newblock ``Supervised learning with quantum-enhanced feature spaces''.
\newblock \href{https://dx.doi.org/10.1038/s41586-019-0980-2}{{Nature} {\bf 567}, 209–212}~(2019).

\bibitem{Shingu2021}
Yuta Shingu, Yuya Seki, Shohei Watabe, Suguru Endo, Yuichiro Matsuzaki, Shiro Kawabata, Tetsuro Nikuni, and Hideaki Hakoshima.
\newblock ``{Boltzmann machine learning with a variational quantum algorithm}''.
\newblock \href{https://dx.doi.org/10.1103/PhysRevA.104.032413}{Physical Review A {\bf 104}, 032413}~(2021).

\bibitem{qchem_Smart2021}
Scott~E. Smart and David~A. Mazziotti.
\newblock ``{Quantum Solver of Contracted Eigenvalue Equations for Scalable Molecular Simulations on Quantum Computing Devices}''.
\newblock \href{https://dx.doi.org/10.1103/physrevlett.126.070504}{{Physical Review Letters} {\bf 126}, 070504}~(2021).

\bibitem{qchem_Gao2021}
Qi~Gao, Hajime Nakamura, Tanvi~P. Gujarati, Gavin~O. Jones, Julia~E. Rice, Stephen~P. Wood, Marco Pistoia, Jeannette~M. Garcia, and Naoki Yamamoto.
\newblock ``{Computational Investigations of the Lithium Superoxide Dimer Rearrangement on Noisy Quantum Devices}''.
\newblock \href{https://dx.doi.org/10.1021/acs.jpca.0c09530}{{The Journal of Physical Chemistry A} {\bf 125}, 1827–1836}~(2021).

\bibitem{Jones2019}
Tyson Jones, Suguru Endo, Sam McArdle, Xiao Yuan, and Simon~C. Benjamin.
\newblock ``{Variational quantum algorithms for discovering Hamiltonian spectra}''.
\newblock \href{https://dx.doi.org/10.1103/PhysRevA.99.062304}{Physical Review A {\bf 99}, 062304}~(2019).

\bibitem{Delgado2021}
Alain Delgado, Juan~Miguel Arrazola, Soran Jahangiri, Zeyue Niu, Josh Izaac, Chase Roberts, and Nathan Killoran.
\newblock ``{Variational quantum algorithm for molecular geometry optimization}''.
\newblock \href{https://dx.doi.org/10.1103/PhysRevA.104.052402}{Physical Review A {\bf 104}, 052402}~(2021).

\bibitem{Abbas2024}
Amira Abbas, Andris Ambainis, Brandon Augustino, Andreas B\"{a}rtschi, Harry Buhrman, Carleton Coffrin, Giorgio Cortiana, Vedran Dunjko, Daniel~J. Egger, Bruce~G. Elmegreen, Nicola Franco, Filippo Fratini, Bryce Fuller, Julien Gacon, Constantin Gonciulea, Sander Gribling, Swati Gupta, Stuart Hadfield, Raoul Heese, Gerhard Kircher, Thomas Kleinert, Thorsten Koch, Georgios Korpas, Steve Lenk, Jakub Marecek, Vanio Markov, Guglielmo Mazzola, Stefano Mensa, Naeimeh Mohseni, Giacomo Nannicini, Corey O’Meara, Elena~Peña Tapia, Sebastian Pokutta, Manuel Proissl, Patrick Rebentrost, Emre Sahin, Benjamin C.~B. Symons, Sabine Tornow, Víctor Valls, Stefan Woerner, Mira~L. Wolf-Bauwens, Jon Yard, Sheir Yarkoni, Dirk Zechiel, Sergiy Zhuk, and Christa Zoufal.
\newblock ``Challenges and opportunities in quantum optimization''.
\newblock \href{https://dx.doi.org/10.1038/s42254-024-00770-9}{Nature Reviews Physics {\bf 6}, 718–735}~(2024).

\bibitem{ibm}
``{IBM Quantum Platform}''.
\newblock \url{https://quantum.ibm.com/}.
\newblock Accessed: 2024.02.25.

\bibitem{braket}
``{Braket}''.
\newblock \url{https://aws.amazon.com/braket}.
\newblock Accessed: 2024.02.25.

\bibitem{a4a_paci}
{Airlines for America}.
\newblock ``{A4A Passenger Airline Cost Index}''.
\newblock \url{https://www.airlines.org/dataset/ a4a-quarterly-passenger-airline-cost-index-u-s-passenger-airlines/}~(2024).
\newblock Accessed: 2024.02.15.

\bibitem{a4a_state}
{Airlines for America}.
\newblock ``{State of U.S. Aviation}''.
\newblock \url{https://www.airlines.org/dataset/state-of-us-aviation/}~(2024).
\newblock Accessed: 2024.02.15.

\bibitem{deveci2018evolutionary}
Muhammet Deveci and Nihan Çetin Demirel.
\newblock ``Evolutionary algorithms for solving the airline crew pairing problem''.
\newblock \href{https://dx.doi.org/10.1016/j.cie.2017.11.022}{Computers \& Industrial Engineering {\bf 115}, 389--406}~(2018).

\bibitem{muter2013solving}
İbrahim Muter, Ş. İlker Birbil, Kerem B\"{u}lb\"{u}l, G\"{u}ven\c{c} Şahin, H\"{u}sn\"{u} Yenig\"{u}n, Duygu Taş, and Dilek T\"{u}z\"{u}n.
\newblock ``Solving a robust airline crew pairing problem with column generation''.
\newblock \href{https://dx.doi.org/10.1016/j.cor.2010.11.005}{Computers \& Operations Research {\bf 40}, 815–830}~(2013).

\bibitem{borndorfer2006column}
Ralf Bornd{\"o}rfer, Uwe Schelten, Thomas Schlechte, and Steffen Weider.
\newblock ``A column generation approach to airline crew scheduling''.
\newblock In Hans-Dietrich Haasis, Herbert Kopfer, and J{\"o}rn Sch{\"o}nberger, editors, Operations Research Proceedings 2005.
\newblock \href{https://dx.doi.org/10.1007/3-540-32539-5_54}{Pages 343--348}.
\newblock Berlin, Heidelberg~(2006). Springer Berlin Heidelberg.

\bibitem{pip}
``{The PyPA recommended tool for installing Python packages}''.
\newblock \url{https://pypi.org/project/pip/}.
\newblock Accessed: 2025.03.26.

\bibitem{Galda2021}
Alexey Galda, Xiaoyuan Liu, Danylo Lykov, Yuri Alexeev, and Ilya Safro.
\newblock ``{Transferability of optimal QAOA parameters between random graphs}''.
\newblock In 2021 IEEE International Conference on Quantum Computing and Engineering (QCE).
\newblock \href{https://dx.doi.org/10.1109/QCE52317.2021.00034}{Pages 171--180}.
\newblock ~(2021).

\bibitem{Lee2021}
Xinwei Lee, Yoshiyuki Saito, Dongsheng Cai, and Nobuyoshi Asai.
\newblock ``{Parameters Fixing Strategy for Quantum Approximate Optimization Algorithm}''.
\newblock In 2021 IEEE International Conference on Quantum Computing and Engineering (QCE).
\newblock \href{https://dx.doi.org/10.1109/QCE52317.2021.00016}{Pages 10--16}.
\newblock ~(2021).

\bibitem{pydantic}
``{Data validation using Python type hints}''.
\newblock \url{https://pypi.org/project/pydantic/}.
\newblock Accessed: 2025.03.26.

\bibitem{Date2021}
Prasanna Date, Davis Arthur, and Lauren Pusey-Nazzaro.
\newblock ``{QUBO formulations for training machine learning models}''.
\newblock \href{https://dx.doi.org/10.1038/s41598-021-89461-4}{Scientific Reports {\bf 11}, 10029}~(2021).

\bibitem{McCollum2021}
Joey McCollum and Thomas Krauss.
\newblock ``{QUBO formulations of the longest path problem}''.
\newblock \href{https://dx.doi.org/https://doi.org/10.1016/j.tcs.2021.02.021}{Theoretical Computer Science {\bf 863}, 86--101}~(2021).

\bibitem{Jun2024}
Kyungtaek Jun.
\newblock ``{QUBO formulations for a system of linear equations}''.
\newblock \href{https://dx.doi.org/https://doi.org/10.1016/j.rico.2024.100380}{Results in Control and Optimization {\bf 14}, 100380}~(2024).

\bibitem{lucas_2014}
Andrew Lucas.
\newblock ``Ising formulations of many {NP} problems''.
\newblock \href{https://dx.doi.org/10.3389/fphy.2014.00005}{Frontiers in Physics {\bf 2}, 5}~(2014).

\bibitem{dataset_Kasirzadeh2017}
Atoosa Kasirzadeh, Mohammed Saddoune, and Fran\c{c}ois Soumis.
\newblock ``{Airline crew scheduling: models, algorithms, and data sets}''.
\newblock \href{https://dx.doi.org/10.1007/s13676-015-0080-x}{EURO Journal on Transportation and Logistics {\bf 6}, 111–137}~(2017).

\bibitem{https://doi.org/10.48550/arxiv.2109.11455}
Rebekah Herrman, Phillip~C. Lotshaw, James Ostrowski, Travis~S. Humble, and George Siopsis.
\newblock ``{Multi-angle Quantum Approximate Optimization Algorithm}''~(2021).

\bibitem{Vikstl2020}
Pontus Vikstål, Mattias Gr\"{o}nkvist, Marika Svensson, Martin Andersson, G\"{o}ran Johansson, and Giulia Ferrini.
\newblock ``{Applying the Quantum Approximate Optimization Algorithm to the Tail-Assignment Problem}''.
\newblock \href{https://dx.doi.org/10.1103/physrevapplied.14.034009}{{Physical Review Applied} {\bf 14}, 034009}~(2020).

\bibitem{https://doi.org/10.48550/arxiv.2205.01192}
Michelle Chalupnik, Hans Melo, Yuri Alexeev, and Alexey Galda.
\newblock ``{Augmenting {QAOA} Ansatz with Multiparameter Problem-Independent Layer}''~(2022).

\bibitem{Vijendran2024}
V.~Vijendran, Aritra Das, Dax~Enshan Koh, Syed~M. Assad, and Ping~Koy Lam.
\newblock ``{An expressive ansatz for low-depth quantum approximate optimisation}''.
\newblock \href{https://dx.doi.org/10.1088/2058-9565/ad200a}{Quantum Science and Technology {\bf 9}, 025010}~(2024).

\bibitem{Powell1994}
M.~J.~D. Powell.
\newblock ``{A Direct Search Optimization Method That Models the Objective and Constraint Functions by Linear Interpolation}''.
\newblock \href{https://dx.doi.org/10.1007/978-94-015-8330-5_4}{Page 51–67}.
\newblock Springer Netherlands. ~(1994).

\bibitem{Spall1992}
J.C. Spall.
\newblock ``{Multivariate stochastic approximation using a simultaneous perturbation gradient approximation}''.
\newblock \href{https://dx.doi.org/10.1109/9.119632}{IEEE Transactions on Automatic Control {\bf 37}, 332–341}~(1992).

\bibitem{Bhatnagar2013}
S.~Bhatnagar, H.L. Prasad, and L.A. Prashanth.
\newblock ``{Stochastic Recursive Algorithms for Optimization: Simultaneous Perturbation Methods}''.
\newblock \href{https://dx.doi.org/10.1007/978-1-4471-4285-0}{Pages 41--44}.
\newblock Springer London. ~(2013).

\bibitem{Acampora2023}
Giovanni Acampora, Angela Chiatto, and Autilia Vitiello.
\newblock ``{Genetic algorithms as classical optimizer for the Quantum Approximate Optimization Algorithm}''.
\newblock \href{https://dx.doi.org/10.1016/j.asoc.2023.110296}{Applied Soft Computing {\bf 142}, 110296}~(2023).

\end{thebibliography}

\newpage
\onecolumn
\appendix

\section{QUBO in the Ising model}
\label{appendix:QUBO2Ising}

Here we show how the QUBO formulation in \eqref{eq-qubo} can be rearranged to describe a Hamiltonian in the Ising model.

As seen in the work of Vikst\aa l et al.~\cite{Vikstl2020}, by replacing the variables $x_i\in\{0,1\}$ with spin variables $\omega_i\in\{-1,1\}$ in \eqref{eq-qubo}, the value of our objective function is equivalent to the energy of the spin configuration $\omega$ given by the Hamiltonian function:
\begin{equation}\label{eq:energy}
    \mathcal{H}(\omega)=D\sum_{i=1}^{n} \left( 1- \sum_{j=1}^{m}b_{ij}\frac{\omega_j+1}{2} \right)^2 + \sum_{j=1}^{m}c_j\frac{\omega_j+1}{2}\enspace.
\end{equation}

Notice that expanding the squares and collecting terms yields the following:
\begin{equation}
\begin{split}
    \mathcal{H}(\omega)={}&\frac{D}{4} \sum_{i=1}^{n}\sum_{j=1}^{m}\sum_{j'=1}^{m}b_{ij}b_{ij'}\omega_j\omega_{j'}\\
    &+ \frac{D}{2} \sum_{i=1}^{n}\sum_{j=1}^{m} b_{ij}\omega_j \bigg(\sum_{j'=1}^{m}b_{ij'} - 2 \bigg) + \sum_{j=1}^{m}\frac{\omega_jc_j}{2}\\
    &+ \frac{D}{4}\sum_{i=1}^{n} \left(\sum_{j=1}^{m}b_{ij} - 2 \right)^2 + \sum_{j=1}^{m}\frac{c_j}{2}.
\end{split}
\end{equation}

By defining 
\begin{align}
    J_{jj'} &= \frac{D}{2}\sum_{i=1}^n b_{ij}b_{ij'} \quad & \forall j,j' \in \{1\ldots m\}, \label{coeffs_J} \\
    h_j &= \frac{D}{2}\sum_{i=1}^n b_{ij}\left( \sum_{j'=1}^m b_{ij'}-2 \right) + \frac{c_j}{2} \quad & \forall j \in \{1\ldots m\}, \label{coeffs_h}\\
    const &= \frac{D}{4}\sum_{i=1}^{n}\left( \sum_{j=1}^{m} b_{ij} - 2 \right)^2 + \sum_{j=1}^{m}\frac{c_j}{2}\label{const},
\end{align}
the Hamiltonian function becomes
\begin{equation}\label{eq:ising-const}
     \mathcal{H}(\omega)=1/2\sum_{j=1}^{m}\sum_{j'=1}^{m} J_{jj'}\omega_j\omega_{j'}+\sum_{j=1}^{m}h_j\omega_j + const\enspace.
\end{equation} 
Since $J$ is symmetric and for all $i$, $J_{ii}\omega_i\omega_i = J_{ii}(\pm1)^2 = J_{ii}$ we can further rewrite it as:
\begin{equation}
    \mathcal{H}(\omega)=\sum_{\substack{j,j'=1 \\ j < j'}}^{m} J_{jj'}\omega_j\omega_{j'}+\sum_{j=1}^{m}h_j\omega_j + const + \sum_{j=1}^{m}J_{jj}/2\enspace.
\end{equation}
From this point, we will leave out the constant terms $const +\sum_{j=1}^{m}J_{jj}/2$ from the formulation as they do not affect which spin configurations have the lowest energy and can be added back later if we wish to know the exact energy of a given configuration. With this, we have arrived at a Hamiltonian\footnote{Note that this operator is exponential in size but sparse, and we can write it using polynomially many terms} similar in form to that in \eqref{cost_hamiltonian} repeated here: 
\begin{equation*}
    \mathcal{H}_C=\sum_{\substack{j,j'=1 \\ j < j'}}^{m} J_{jj'}\sigma^z_j \sigma^{z}_{j'}+\sum_{j=1}^{m}h_j \sigma^z_j\enspace.
\end{equation*}

\section{Airline Crew Pairing}
\subsection{Pairing generation}\label{generation}
To generate the pairings, first we take the flight legs of each day and generate the valid duties from them. In the sequence, consecutive flight legs must come after one another in time. The arrival airport of the former must also match the departure airport of the latter. Additional validation rules may be applied according to the configuration provided.

With the multiple duties for each day generated, these are in turn, joined into pairings.
Pairings must start and end at the same airport, the home base of the crew.
Here, just like with duties, additional validation rules may also be applied. 

This full enumeration scheme on the pairings is a straightforward solution suitable for instance sizes that current quantum computing devices can handle when solving the problem. 

\subsection{Rules}\label{rules}
In our implementation module, we use a rule set inspired by rules that may appear in real-world instances of ACP. The rules are parametric, making it easier to match real-life scenarios by only adjusting the configurations.
\begin{enumerate}  
\item The rule \texttt{max flights} defines the number of flights a duty can contain at most, by default $4$.
\item The \texttt{min connect} rule defines a minimum connection time between flight legs in a duty measured in minutes, $30$ by default.
\item The \texttt{max duration duty time} caps the duty time, that is the time between the first departure and last arrival of a duty, to the given number of hours, by default $12$.
\item The rule \texttt{max duties} ensures that each pairing can only contain a certain number of duties at most, by default $5$.
\item The rule \texttt{min rest} guarantees that between every duty, a certain number of hours ($9.5$ by default) must pass, allowing the crews to rest.
\item The \texttt{max pairing duration} rule ensures that no pairing can last longer than the given number of days, $4$ by default.
\item Finally, the \texttt{max work time} rule sets a cap on the hours worked by crews in each duty, $8$ by default.
\end{enumerate} 

\subsection{Cost model}\label{cost}
The costs of the pairings are determined using \texttt{CostModel} plugins. The example, \texttt{acp-cost-example}, penalizes nights spent not at the home base of the crew and time spent working before 5 AM and after 8 PM.

\newpage
\section{Alternative ans\"atze choices}\label{appendix:ansatze}
Since the introduction of the QAOA, there have been many explorations into different circuit designs that utilize more parameters or different mixer Hamiltonians.

\subsection{ma-QAOA}
The \emph{Multi-angle QAOA} (ma-QAOA) ansatz \cite{https://doi.org/10.48550/arxiv.2109.11455} assigns a variational parameter to each summand in the problem Hamiltonian $\mathcal{H}_C$ and mixing Hamiltonian $\mathcal{H}_M$, generalizing the original formulation where the cost and mixer Hamiltonians had one variational parameter each. Instead of each layer $i$, $i = 1\dots p$, having $2$ variational parameters, $\gamma_i$ and $\beta_i$, we now have parameters $\hat{\gamma}_i \in \mathbb{R}^{m\cross m}$, $\hat{\theta}_i \in \mathbb{R}^{m}$ and $\hat{\beta}_i \in \mathbb{R}^m$ such that:
\begin{equation}\label{ma-qaoa-uc}
    U = U_M(\hat{\beta}_p)U_C(\hat{\gamma}_p, \hat{\theta}_p)...U_M(\hat{\beta}_1)U_C(\hat{\gamma}_1, \hat{\theta}_1),
\end{equation} where
\begin{align}
       U_C(\gamma, \theta) &= \prod_{j'}^m\prod_{\substack{j=1 \\ j < j'}}^{m}R_{ZZ}(2\gamma_{jj'} J_{jj'})_{jj'}\prod_j^mR_{Z}(2\theta_{j} h_j)_j\textrm{, and}\\
       U_M(\beta) &= \prod^m_jR_{X}(2\beta_{j})_j.
\end{align}
The ma-QAOA ansatz is implemented in the \texttt{maqaoa-ansatz} plugin.

\subsection{QAOA+}
The QAOA+ ansatz~\cite{https://doi.org/10.48550/arxiv.2205.01192} augments the original QAOA ansatz by adding a problem-independent layer of parametric quantum circuits, allowing it to outperform the original QAOA at shallower depths and even outperform the ma-QAOA in some cases. The circuit implements the unitary transformation
\begin{equation}
    U = U_+(\hat{\nu}, \hat{\mu})U_M(\hat{\beta}_p)U_C(\hat{\gamma}_p)...U_M(\hat{\beta}_1)U_C(\hat{\gamma}_1),
\end{equation}
where $\hat{\nu}, \hat{\mu} \in \mathbb{R}^m$ and $U_+$ is the problem independent unitary transformation,
\begin{equation}
    U_+(\nu, \mu) =
    \prod^{m}_{j=2}R_{ZZ}(\nu_j)_{j,j-1}
    \prod^{m}_{j=1}R_X(\mu_j)_j\enspace.
\end{equation}
The ansatz is available through the \texttt{qaoa-plus-ansatz} plugin.

\subsection{XQAOA}
Implemented as the \texttt{xqaoa-ansatz} plugin, the \emph{eXpressive QAOA} (XQAOA) ansatz~\cite{Vijendran2024} is a generalization of ma-QAOA. It extends the mixer Hamiltonian by adding in Y-rotations, parametrized using a vector $\hat{\alpha}_i \in \mathbb{R}^m$ for each layer $i = 1\dots p$ to increase the expressiveness of the ansatz by being able to express any computational-basis state with appropriate parametrization using the unitary
\begin{equation}
    U = U_M(\hat{\alpha}_p,\hat{\beta}_p)U_C(\hat{\gamma}_p, \hat{\theta}_p)...U_M(\hat{\alpha}_1,\hat{\beta}_1)U_C(\hat{\gamma}_1, \hat{\theta}_1),
\end{equation} where
$U_C$ is similar to \eqref{ma-qaoa-uc} used in ma-QAOA and
\begin{align}
       U_M(\beta,\alpha) &= \prod^m_jR_{Y}(2\alpha_{j})_j\prod^m_jR_{X}(2\beta_{j})_j\enspace.
\end{align}
Table \eqref{num-params} summarizes the number of variational parameters used in these ans\"atze.

\begin{table}
    \centering
    \captionsetup{width=.8\linewidth}
  \caption{Number of parameters for ans\"atze in Appendix \eqref{appendix:ansatze} depending on algorithm depth $p$ and number of variables $m$}
  \label{num-params}
  \begin{tabular}{ccl}
    \toprule
    Ansatz& Number of parameters\\
    \midrule
     QAOA& $2p$\\
     ma-QAOA& $p(m^2+2m)$\\
     QAOA+& $2(p+m)$\\
     XQAOA& $p(m^2+3m)$\\
  \bottomrule
\end{tabular}
\end{table}

\newpage
\section{Optimization strategies for finding optimal parameters}\label{appendix:opt}
This section highlights a few methods to optimize the parameters that are available in the suite.

\subsection{COBYLA}
The COBYLA (Constrained Optimization BY Linear Approximations) algorithm~\cite{Powell1994} operates by iteratively constructing linear approximations of both the objective function and constraints using a simplex of \( n+1 \) points in an \( n \)-dimensional space. At each iteration, it optimizes within a trust region, updating the simplex to improve feasibility and objective value.
This method is usable by appropriately configuring the \texttt{scipy-optimizer} plugin.

\subsection{SPSA}
The \emph{Simultaneous Perturbation Stochastic Approximation} (SPSA)~\cite{Spall1992, Bhatnagar2013} method is a gradient-free optimization algorithm that requires only two evaluations of the cost function $C$ in every iteration. This can help to reduce the costs incurred by executing the quantum subroutines of VQAs. In the SPSA algorithm the $k$th estimate $\hat{\theta}_k \in \mathbb{R}^p$ of the optimal parameter $\theta^* \in \mathbb{R}^p$ is defined as
\begin{equation}
    \hat{\theta}_{k+1} = \hat{\theta}_k - a_k\hat{g}_k(\hat{\theta}_k),
\end{equation}
where $\hat{g}_k(x)$ is the approximate gradient of the cost function at point $x \in \mathbb{R}^p$ and the sequence ${a_k}$ satisfies conditions described by Spall~\cite{Spall1992}.
This method is available in the suite as the \texttt{spsa-optimizer} plugin.

\subsection{Genetic Algorithms}
The optimization landscape of QAOA, especially at large depths, is rather complex, characterized by many local minima and barren plateaus, regions with vanishingly small gradients, which makes finding the optimal parameters increasingly difficult.
The use of genetic algorithms~\cite{Acampora2023} aims to alleviate these concerns by leveraging population-based metaheuristics to manage candidate solutions and starting from a random population search for the optimal solution by applying stochastic operators.

In genetic algorithms, inspired by Darwinian evolution, an initial population of candidate solutions evolves stochastically over generations. Candidate solutions (chromosomes) are strings of numbers (genes), which in our case are the ansatz parameters. Every generation, genetic operators are applied to the population to determine the population of the next generation. This involves selecting the best chromosomes from the population using a fitness function, which means that a chromosome that achieves a lower energy state of the quantum system is deemed fitter.
A configurable genetic algorithm is available through the \texttt{genetic-optimizer} plugin.

\end{document}